\documentclass[reqno]{amsart}
\usepackage{amsmath}
\usepackage{amsfonts}
\usepackage{amssymb}
\usepackage{graphics}
\usepackage{graphicx}
\usepackage{tikz}
\usetikzlibrary{shapes.geometric}
\usetikzlibrary{positioning, calc}
\usepgflibrary{arrows}

\newtheorem{theorem}{Theorem}[section]
\newtheorem{corollary}{Corollary}

\newtheorem{definition}[theorem]{Definition}
\newtheorem{remark}{Remark}

\newtheorem{example}[theorem]{Example}

\DeclareMathOperator{\Hom}{Hom}

\title{On the quotient quantum graph with respect to the regular representation}
\author{G\"{o}khan Mutlu}

\begin{document}

\maketitle

\begin{abstract}
Given a quantum graph $ \Gamma $, a finite symmetry group $ G $ acting on it and a representation $ R $ of $ G $, the quotient quantum graph $ \Gamma /R $ is described and constructed in the literature \cite{ana, bandpargil, parband}. In particular, it was shown that the quotient graph $ \Gamma/\mathbb{C}G $ is isospectral to $ \Gamma $ by using representation theory where $ \mathbb{C}G $ denotes the regular representation of $ G $ \cite{parband}. Further, it was conjectured that $ \Gamma $ can be obtained as a quotient $ \Gamma/\mathbb{C}G $ \cite{parband}. However, proving this by construction of the quotient quantum graphs has remained as an open problem. In this paper, we solve this problem by proving by construction that for a quantum graph $ \Gamma $ and a finite symmetry group $ G $ acting on it, the quotient quantum graph $ \Gamma / \mathbb{C}G $ is not only isospectral but rather identical to $ \Gamma $ for a particular choice of a basis for $ \mathbb{C}G $. Furthermore, we prove that, this result holds for an arbitrary permutation representation of $ G $ with degree $ |G| $, whereas it doesn't hold for a permutation representation of $ G $ with degree greater than $|G|. $ 
\end{abstract}

\section{Introduction}

Quantum graphs enable us to model systems in mathematics, physics, chemistry, and engineering to analyze phenomena such as the free-electron
theory of conjugated molecules, quantum wires, dynamical systems, photonic crystals, thin waveguides, and many other. Besides, in 1997, Kottos and Smilansky \cite{kotsmi} proposed quantum graphs as a model to study quantum chaos and then the research on this field has grown extensively. Recently, there is a substantial interest in this field especially on isospectrality, nodal domains and nodal count, trace formulas, quantum chaos and spectral statistics. The interested reader about quantum graphs is referred to \cite{ber, berkuc, gnusmi,kuc, mug} and references therein.

In 1966, Marc Kac asked the famous question: "Can one hear the shape of a drum?" which means "does the Laplacian on every planar domain with Dirichlet boundary conditions have a unique spectrum?"\cite{kac}. This question triggered the research on isospectrality and inverse problems. Many attempts were made both to construct isospectral but different objects or to recover the shape of an object from its spectrum in order to answer Kac's question. The first fundamental contribution to answer this question was made by Sunada who gave a method for constructing isospectral Riemannian manifolds \cite{sun}. Finally, Gordon et al. answered Kac's question negatively by constructing the first pair of isospectral planar Euclidean domains by extending Sunada's method \cite{gorweb, gorweb2}. We refer the interested reader about isospectrality to \cite{gor, sha2}. 

In 2001, Gutkin and Smilansky rephrased Kac's question for quantum graphs \cite{gutsmi}. This question was answered negatively in the general case \cite{CARL, von}. However, Gutkin and Smilansky \cite{gutsmi} showed that quantum graphs that have rationally independent bond lengths "can be heard" which means their spectrum determines their connectivity matrices and bond lengths uniquely. There were many attempts to create isospectral quantum graphs. In particular, Band et al. constructed a pair of isospectral graphs in \cite{shasmi} and then in \cite{bandpargil, parband} the authors generalized this construction and provided a rigorous recipe to construct quotient quantum graphs by using linear representations. Namely, given a quantum graph $ \Gamma $, a symmetry group $ G $ acting on $ \Gamma $ and a representation $ R $ of $ G $, they constructed another quantum graph $ \Gamma/R $ called a quotient quantum graph. This construction is based on an "encoding scheme" in which one builds the eigenfunctions $ f $ on $ \Gamma/R $ with the  information on the eigenfunctions $ \tilde{f} $ defined on the original graph $ \Gamma $ such that $ \tilde{f} $ transforms according to $ R $ \cite{bandpargil, parband}. Different choices for the fundamental domain of the action of $ G $ on $ \Gamma $ and for the basis of $ R $ yield different quotient graphs $ \Gamma/R $ which are all isospectral to each other. This is a very effective way to create isospectral quantum graphs. Recently, Band et al. revised this construction method for quantum graphs and even applied it to define quotient operators of abstract finite dimensional operators \cite{ana}. 

In \cite{parband}, it was shown that for a quantum graph $ \Gamma $ and a symmetry group $ G $ acting on $ \Gamma $, the quotient graph $ \Gamma/\mathbb{C}G $ is isospectral to $ \Gamma $ where $ \mathbb{C}G $ denotes the regular representation of $ G $. Hence, if we take any basis for the representation $ \mathbb{C}G $ we will obtain a quotient quantum graph $ \Gamma/\mathbb{C}G $ which is isospectral to $ \Gamma $. This is an extremely useful way to create quantum graphs which are isospectral to a given quantum graph. The authors proposed the natural question: "can one obtain $ \Gamma $ by construction as a quotient $ \Gamma/\mathbb{C}G $?" \cite{parband}. In other words, does there exist a basis of $ \mathbb{C}G $ which yields $ \Gamma $? If yes, which particular basis of $ \mathbb{C}G $ gives us $ \Gamma $ when we construct $ \Gamma/\mathbb{C}G $ and how?    

In this paper, we answer the above open question affirmatively. For a quantum graph $ \Gamma $ and a finite symmetry group $ G $ acting on it, we construct the quotient quantum graph $ \Gamma/\mathbb{C}G $ by choosing $ G $ as a basis for $ \mathbb{C}G $ and show that the resulting graph is identical to $ \Gamma $. Moreover, we prove a more general result that $ \Gamma $ can be obtained as a quotient $ \Gamma/\rho $ where $ \rho $ is an arbitrary permutation representation of $ G $ with degree $  |G| $. We prove that if one constructs the quotient graph $ \Gamma/\rho $ by choosing the standard basis of $ \mathbb{C}^{|G|}  $, one gets $ \Gamma $ where $ \rho $ is an arbitrary permutation representation of $ G $ with degree $  |G| $. We also show by a counterexample that this does not hold for a permutation representation of $ G $ with degree greater than $ |G| $.  We use the construction method and notations from \cite{ana} since it is more convenient for the sake of computations. The outline of this paper is as follows. Firstly, we present a brief introduction to provide the necessary notions on quantum graphs, linear representations and the construction of the quotient quantum graphs. Then, we provide an example which motivates our results. Finally, we present the main results and additional examples. 

\section{Preliminaries}      
\subsection{\label{sec21}A brief introduction to quantum graphs}    
Let $ \Gamma $ be a graph with a finite vertex set $ V $ and an edge set $ E $. If we assign a positive length $ l_{e} $ and a direction to each edge $ e\in E $ we get a metric graph. In a metric graph we can identify each edge $ e\in E $ with an interval $ [0,l_{e}] $ of the real line and assign a coordinate $ x_{e} $ to each point on $ \Gamma $ along the interval $ \left[ 0,l_{e}\right]  $. 

A function $ f $ on the metric graph $ \Gamma $ is defined as a vector of functions on each edge of the graph. Hence $$ f=\left( f|_{e_{1}},f|_{e_{2}},\ldots,f|_{e_{|E|}} \right)  $$
where $f|_{e_{i}}: \left[ 0,l_{e_{i}}\right] \longrightarrow \mathbb{C}$ for each $ i=1,2,\ldots,|E| $. In order to define a differential operator called Hamiltonian on $ \Gamma $ we need to introduce the Hilbert space of functions on $ \Gamma $ which is the direct sum of Sobolev spaces on the edges i.e. 
\begin{equation*}
H^{2}\left( \Gamma \right):=\bigoplus_{i=1}^{|E|}H^{2}\left[ 0,l_{e_{i}}\right].  				
\end{equation*}
We are now ready to define the Hamiltonian on the metric graph $ \Gamma $. Mostly, this Hamiltonian is 	
\begin{equation*}
\left( -\bigtriangleup +Q \right) f:=\left( -f_{e_{1}}^{\prime \prime}+Qf_{e_{1}}, -f_{e_{2}}^{\prime \prime}+Qf_{e_{2}},\ldots,-f_{e_{|E|}}^{\prime \prime}+Qf_{e_{|E|}} \right), 
\end{equation*}
where $ f_{e_{i}}=f|_{e_{i}} $ and $ Q $ is a potential.

In addition, we need some boundary conditions on $ \Gamma $ known as vertex conditions which relate the values and the first derivatives of functions on $ \Gamma $. The most common vertex conditions are called Neumann or Neumann-Kirchhoff conditions that are given for the vertex $ v \in V $ as follows:

$ (i) $ $ f $ is continuous on vertex $ v $ which means $ f $ agrees on all edges incident to $ v $, 

$ (ii) $ $ \sum_{e \sim  v}\frac{df}{dx}\left( v\right)=0 $ where the sum is taken over all edges incident to $ v $ and the derivatives are taken in the direction from the vertex to the edge. 

One can collect all the vertex conditions into a matrix in the following way. Let us define the operators 
$ \gamma_{D}, \gamma_{N}:H^{2}\left( \Gamma \right)\longrightarrow \mathbb{C}^{2|E|}  $
\begin{equation*}
\gamma_{D}\left( f \right)= \begin{pmatrix}
f_{e_{1}} \left( 0 \right) \\ f_{e_{1}}\left( l_{e_{1}}\right)\\\vdots\\f_{e_{|E|}}\left( 0 \right)\\ f_{e_{|E|}}\left( l_{e_{|E|}} \right) 	
\end{pmatrix}, \quad \gamma_{N}\left( f\right)= \begin{pmatrix}
f^\prime_{e_{1}}\left( 0\right)\\ -f^\prime_{e_{1}}\left( l_{e_{1}}\right)\\ \vdots \\f^\prime_{e_{|E|}}\left( 0\right)\\ -f^\prime_{e_{|E|}}\left( l_{e_{|E|}}\right) 	
\end{pmatrix}
\end{equation*}
for all $ f \in H^{2}\left( \Gamma \right)  $ where $ f_{e_{i}}=f|_{e_{i}} $. Then, we can represent all vertex conditions with a single equation
\begin{equation}\label{q1}
A\gamma_{D}\left( f\right)+B\gamma_{N}\left( f\right)=0,
\end{equation} 
where $ A $ and $ B $ are $ 2|E|\times 2|E| $ matrices.
The Hamiltonian is defined for all $ f \in H^{2}\left( \Gamma \right)  $ such that (\ref{q1}) is satisfied. This operator is selfadjoint iff $ Rank \left( A \mid B \right)=|E|  $ and $ AB^{*} $ is selfadjoint \cite{kos} where $ A^{*} $ denotes the adjoint of the matrix $ A $. The metric graph $ \Gamma $ together with the Hamiltonian and vertex conditions given by \eqref{q1} is called a quantum graph. To sum it up, a quantum graph is a metric graph with an operator (Hamiltonian) defined on it and the vertex conditions given at each vertex. More information about quantum graphs can be found in \cite{ber, berkuc, gnusmi,kuc, mug}. 
\subsection{\label{sec22}Some basic notions on linear representations}
Let $ G $ be a group and $\mathcal{V} $ be a finite dimensional complex vector space. A (linear) representation of $ G $ is a group homomorphism $ R:G\longrightarrow GL\left(\mathcal{V} \right) $ where $ GL\left( \mathcal{V} \right) $ is the group of all automorphisms of $ \mathcal{V} $. In this case, $ \mathcal{V} $ is called the carrier-space of the representation and often denoted by $ \mathcal{V}_{R} $. The degree of the representation $ R $ is defined as the number $ \dim \mathcal{V} $.

Let $ R:G\longrightarrow GL\left( \mathcal{V} \right) $ be a representation of $ G $. Then, $ \mathcal{V} $ is endowed with the linear group action defined by $ gv:=R\left( g \right)v $ for all $ v\in \mathcal{V} $ and hence $ \mathcal{V} $ is a $ \mathbb{C}G $-module. Conversely, if $ \mathcal{V} $ is a $ \mathbb{C}G $-module i.e., $ G $ acts linearly on $ \mathcal{V} $, define $ R:G\longrightarrow GL\left( \mathcal{V} \right) $ such that for all $g\in G $, 
\begin{equation*}
R\left( g \right):\mathcal{V}\longrightarrow \mathcal{V}, \quad R\left( g \right)v:=gv. 
\end{equation*}
Then, $ R $ is a representation of $ G $. Thus, there is a one-to-one correspondence between the representations of $ G $ and $ \mathbb{C}G $-modules. We can use the definitions interchangeably and refer to both the representation $ R $ and the $ \mathbb{C}G $-module $ \mathcal{V}_{R} $ as the representation.  

Let $ X $ be a finite set and the finite group $ G $ acts on the left on $ X $. Let $ \mathcal{V} $ be the vector space generated by the basis $ \{e_{x}: x \in X \} $. Then, $ G $ acts linearly on $ \mathcal{V} $ by permuting the basis vectors i.e., 
\begin{equation*}
g\left( \sum a_{x}e_{x} \right)=\sum a_{x}e_{gx}.
\end{equation*}
In other words, $ \rho: G\longrightarrow GL_{n}\left( \mathbb{C} \right) $ is a (linear) representation such that each $ g \in G $ is mapped to the corresponding permutation matrix $ \rho\left( g\right)  $, where $ n=|X| $. $ \rho $ is called a (linear) permutation representation of $ G $. If we take $ X=G $ and the left action as the left multiplication by a group element, then by the above definition $ \rho_{G}: G\longrightarrow GL_{|G|}\left( \mathbb{C} \right) $ is a (linear) permutation representation of $ G $ called the regular representation. Clearly, degree of $ \rho_{G} $ is $ |G| $ and the carrier-space of $ \rho_{G} $ is the group algebra $ \mathbb{C}G $ considered as a complex vector space. Note that the set of all group elements is a basis of this space. Assume elements of $ G $ are ordered $ G=\{g_1,g_2,\ldots,g_{|G|}\} $ to establish such a basis. Then, we can represent the regular representation
\begin{equation*}
\left[ \rho(g) \right]_{ij}=\delta \left( gg_{j},g_{i}\right), \ i,j=1,2,\ldots,|G|,  
\end{equation*}
for each $ g \in G $ where $\delta$ is the Kronecker delta. With abuse of notation we refer to both $ \mathbb{C}G $ and the representation $ \rho_{G} $ as the regular representation of $ G $. 

Let $ G $ be a finite group with two representations $ \rho: G\longrightarrow~GL\left( \mathcal{V}_{\rho} \right) $ and $ \sigma: G\longrightarrow GL\left( \mathcal{V}_{\sigma} \right) $ with degrees $ r $ and $ p $ respectively. The vector space homomorphism $ \phi:\mathcal{V}_{\rho}\longrightarrow \mathcal{V}_{\sigma} $ is called an intertwiner if it satisfies the equality
\begin{equation}\label{int}
\sigma\left( g\right)\phi=\phi\rho\left( g\right), \quad \forall g\in G. 
\end{equation}  
In this case, we say that an $ r $-tuple of vectors $ f_{k}\in \mathbb{C}^{p} $, $ k=1,\ldots,r $ transforms under $ \sigma $ according to $ \rho $ if
\begin{equation*}
\sigma\left( g\right)f_{k}=\sum_{i=1}^{r}\left[\rho\left( g\right) \right]_{ik}f_{i}, \quad \forall g\in G, \quad k=1,\ldots,r.  
\end{equation*}
The set of all intertwiners is a vector space denoted by $  \Hom_{G}\left(\mathcal{V}_{\rho},\mathcal{V}_{\sigma}\right) $. More information on the (linear) representations can be found in \cite{ful}.      

\subsection{\label{sec23}A construction method for quotient quantum graphs}	
There are two construction methods in the literature for obtaining quotient quantum graphs presented in \cite{bandpargil, parband} and in \cite{ana}. The first method is based on an "encoding scheme" which we will not go in detail and rather use the construction method and same notations from \cite{ana} since it is simpler in means of calculation and is a more compact and explicit construction. We summarize the method in \cite{ana} for the sake of completeness. 

Assume $ \Gamma $ is a quantum graph with a finite vertex set $ V $ and an edge set $ E $ and the Hamiltonian is defined as $ -\bigtriangleup +Q $ for some potential $ Q $. Assume also $ G $ is a finite symmetry group acting on $ \Gamma $ and the vertex conditions are given by the equation 
\begin{equation}\label{204}
A\gamma_{D}\left( f\right)+B\gamma_{N}\left( f\right)=0, 
\end{equation} 
as described in Section~\ref{sec21}. We need to specify the properties of this action (symmetry) in order to obtain the quotient graph. For this reason, firstly we require the action maps vertices to vertices that is $gv\in V $, for all $  v\in V, g\in G $ and the action preserves the graph's metric structure, i.e.

$ (i) $  $ l_{ge}= l_{e}, \quad \forall e\in E, \quad \forall g\in G $ (it preserves the edge lengths),

$ (ii) $ $ ge $ and $ e $ have the same direction for all $e\in E$, $g\in G $ (it preserves the edge directions).

The above conditions can be achieved by assuming (i) holds and in addition assuming without loss of generality, an edge is never mapped to its reversal (If such an edge exists, we can add a dummy vertex possessing Neumann vertex conditions in the middle of the edge. By doing so, the action still maps vertices to vertices). With this assumption, we can always assign a direction to each edge such that the action preserves the edge directions. We can regard the action of $ G $ on $ \Gamma $ which preserves the metric structure of the graph as a permutation representation of $ G $ with degree $ |E| $ since the group action permutes the edges. Namely, there exists a representation $ \pi:G\longrightarrow GL_{|E|} \left( \mathbb{C} \right)  $ such that $ \pi\left( g\right) $ is a permutation matrix for each $ g\in G $.

Secondly, we require the action respects the quantum graph structure, namely:

$ (iii) $ $ Q_{ge}= Q_{e}, \quad \forall e\in E, \quad \forall g\in G $ (the action preserves the graph's potential),

$ (iv) $ (the action) $ \pi $ preserves the vertex conditions, namely;
\begin{equation*}
A\gamma_{D} \left( f \right) + B\gamma_{N} \left( f \right)=0 \Leftrightarrow A\hat{\pi} \left( g \right) \gamma_{D}\left( f\right)+B\hat{\pi}\left( g \right)\gamma_{N}\left( f\right)=0
\end{equation*}
for all $ g\in G $ with $ \hat{\pi}\left( g \right)=\pi\left( g \right)\otimes I_{2} $ where $ I_{n} $ denotes the $ n \times n $ unit matrix.

In this case, $ \Gamma $ is called a $ \pi $-symmetric quantum graph \cite{ana}. The matrices $ A $ and $ B $ in Equation (\ref{204}) are called $\hat{\pi}$-symmetric if 
\begin{equation*}
\hat{\pi}\left( g \right)^{\ast}A\hat{\pi}\left( g \right)=A,\ \ \hat{\pi}\left( g \right)^{\ast}B\hat{\pi}\left( g \right)=B.
\end{equation*}

Note that $ A $ and $ B $ are not $\hat{\pi}$-symmetric in general, but there exist always matrices $ \tilde{A}=\left( A+iB\right)^{-1}A $ and $ \tilde{B}=\left( A+iB\right)^{-1}B $ which are $\hat{\pi}$-symmetric and give the equivalent vertex conditions (see Lemma 6.3 in \cite{ana}). Hence we can assume from now on that the matrices $ A $ and $ B $ in Equation (\ref{204}) are in fact $\hat{\pi}$-symmetric.
\begin{definition}\label{def}
	Let $ \pi:G \longrightarrow GL_{|E|} \left( \mathbb{C} \right)$ be a permutation representation of the finite group $ G $ acting on the quantum graph $ \Gamma $ such that $ \Gamma $ is $ \pi $-symmetric. Assume the vertex conditions are given by (\ref{204}). Let $ \rho:G\longrightarrow GL_{r}\left( \mathbb{C} \right) $ be another representation of $ G $ with degree $ r $. Assume $ \Theta $ is an $ r|E|\times d $ matrix whose columns form an orthonormal basis for the kernel space associated to representation $ \rho $ 
	\begin{equation*}
	K_{G}\left( \rho, \pi \right):=\bigcap_{g \in G} \ker \left[ I_{r} \otimes \pi\left( g\right) - \rho \left( g \right)^{T}\otimes I_{|E|} \right], 
	\end{equation*}
	where $ d:=\dim K_{G}\left( \rho, \pi \right)=\dim \Hom_{G}\left(\mathcal{V}_{\rho},\mathcal{V}_{\pi}\right)  $ and $ A^{T} $ denotes the transpose of the matrix $ A $. Let us choose a fundamental domain $ D=\{e_{1},e_{2},\ldots,e_{|D|}\} $ for the action of $ G $ by selecting only one edge from each orbit $ O_{j}=\{e_{k} \in E : \exists g\in G, e_{k}=ge_{j} \} $. $ d_{i} $ is the number of the copies of the edge $ e_{i}  $ in the quotient graph and is given by 
	\begin{equation}\label{211}
	d_{i}=\dim \bigcap _{g \in G_{e_{i}}} \ker \left[ I_{r}-\rho\left( g \right)^{T} \right], 
	\end{equation}
	where $ G_{e_{i}} $ denotes the stabilizer of $ e_{i} $. Thus, if the edge $ e_{i} $ has a trivial stabilizer then we will have $ r $ copies of the edge in the quotient but for the other vertices this number reduces to $ d_{i} $ (see Remark 2.9 \cite{ana}). The quotient quantum graph $ \Gamma_{\rho} $ is formed from the edges $ \{e_{i,j}\}_{i \in D, j=1,\ldots,d_{i}} $ with corresponding lengths $ \{l_{e_{i}}\}_{i \in D } $ where $ d=\sum_{i \in D}d_{i} $. The Hamiltonian on this quotient graph is defined by the differential expression $ -d^{2}/dx^{2}+Q_{e_{i}} $ on the edges $ e_{i,j} $. The vertex conditions are given by the equations
	\begin{equation}\label{207}
	A_{\rho}:=\hat{\Theta}^{\ast} \left[ I_{r}\otimes A \right]\hat{\Theta},\quad B_{\rho}:=\hat{\Theta}^{\ast} \left[ I_{r}\otimes B \right] \hat{\Theta},  
	\end{equation}
	where $ \hat{\Theta}=\Theta \otimes I_{2} $. 
\end{definition}

\begin{remark}
	Note that above definition of the quotient quantum graph is also a quotient in the sense of \cite{bandpargil, parband}, but the converse is not true in general. To make two definitions equivalent one needs to allow the matrices $ \Theta $ whose columns are not orthonormal (see Remark 2.8 \cite{ana}). Above construction of the quotient quantum graphs guarantees the selfadjointness of the resulting graph on the contrary to \cite{bandpargil, parband}. Note also that this construction is based on the matrix $ \Theta $ and hence different choices of $ \Theta $ yield different but isospectral quantum graphs. We used a slightly modified but equivalent construction of the quotient quantum graph, namely; we don't use $ K_{G}\left( \rho, \pi \right) $ in order to obtain $ \Theta $ but equivalently find a basis for the space of intertwiners $ \Hom_{G}\left(\mathcal{V}_{\rho},\mathcal{V}_{\pi}\right) $ (which is shown to be isomorphic to the kernel space (see Lemma 2.5 in \cite{ana})) and then apply the vectorization map to elements of this basis to obtain $ \Theta $.  
\end{remark}    

\section{\label{sec3}Statement of the problem and an example}
In this section, we present the following open problem proposed in \cite{parband} and provide an example to motivate our solution to this problem. In \cite{parband}, the authors proved that for a quantum graph $ \Gamma $ and a symmetry group $ G $ acting on $ \Gamma $, the quotient graph $ \Gamma/\mathbb{C}G $ is isospectral to $ \Gamma $ where $ \mathbb{C}G $ denotes the regular representation of $ G $ (see Proposition 2 \cite{parband}). Further, the authors conjectured that $ \Gamma $ can be obtained as a quotient $ \Gamma/\mathbb{C}G $ \cite{parband}. We try to prove this conjecture which can be rephrased as "does there exist a basis of $ \mathbb{C}G $ which yields $ \Gamma $? If yes, which particular basis of $ \mathbb{C}G $ gives us $ \Gamma $ when we construct $ \Gamma/\mathbb{C}G $ and how?" 
\begin{example} \label{example1}
	Let $\Gamma$ be a star graph with three edges, one edge of length $ l_{1} $ and two edges with equal length $ l_{2} $ (see Figure 1). All vertices are equipped with Neumann conditions. We take $G=C_2=\{I, R\}$, the cyclic group of order 2 generated by $ R $ as a symmetry group of $\Gamma$ where $ R $ is the reflection which exchanges $ e_{2}\leftrightarrow e_{3} $. Clearly, the graph is $\pi$-symmetric, where the representation $\pi$ is 
	$$
	\pi(I)=I_3, \  \   \pi(R)= \left(
	{\begin{array}{rrr}
		1 & 0 & 0 \\
		0 & 0 & 1 \\
		0 & 1 & 0
		\end{array}}
	\right). 
	$$ 
	\begin{figure} [h]
		\centering	
		\begin{tikzpicture}
		\coordinate [label=right:0] (0) at (0,0);
		\coordinate [label=left:1] (1) at (-1.5,0);
		\coordinate [label=right:2] (2) at (1,1);
		\coordinate [label=right:3] (3) at (1,-1);
		\draw (0)--(1);
		\draw (0)--(2);
		\draw (0)--(3);
		
		\draw [-stealth] (0,0)--(-0.5,0);
		\draw [-stealth] (0,0)--(0.5,0.5);
		\draw [-stealth] (0,0)--(0.5,-0.5);
		
		\draw[fill] (0,0) circle [radius=0.1];
		\draw[fill] (-1.5,0) circle [radius=0.1];
		\draw[fill] (1,1) circle [radius=0.1];
		\draw[fill] (1,-1) circle [radius=0.1];
		
		\coordinate[label=below:$l_1$](01) at ($ (0)!.5!(1) $);
		\coordinate[label=right:$l_2$] (03) at ($ (0)!.5!(3) $);
		\coordinate[label=right:$l_2$](02) at ($ (0)!.5!(2) $);
		\coordinate[label=above:$e_1$](01) at ($ (0)!.5!(1) $);
		\coordinate[label=left:$e_2$](02) at ($ (0)!.5!(2) $);
		\coordinate[label=left:$e_3$](03) at ($ (0)!.5!(3) $);
		\end{tikzpicture}
		\caption{A star graph with $ C_{2} $ symmetry. }
	\end{figure}
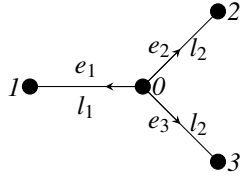
	The vertex conditions of the graph may be given by (see Section~\ref{sec21}) 
	\begin{equation}\label{350}
	A= \left( \begin{matrix} 
	1 & 0 & -1 & 0 & 0 & 0\\
	0 & 0 & 1 & 0 & -1 & 0\\
	0 & 0 & 0 & 0 & 0 & 0 \\
	0 & 0 & 0 & 0 & 0 & 0 \\
	0 & 0 & 0 & 0 & 0 & 0 \\
	0 & 0 & 0 & 0 & 0 & 0
	\end{matrix}
	\right), \ \ 
	B= \left( \begin{matrix} 
	0 & 0 & 0 & 0 & 0 & 0\\
	0 & 0 & 0 & 0 & 0 & 0\\
	1 & 0 & 1 & 0 & 1 & 0 \\
	0 & 1 & 0 & 0 & 0 & 0 \\
	0 & 0 & 0 & 1 & 0 & 0 \\
	0 & 0 & 0 & 0 & 0 & 1
	\end{matrix}
	\right).
	\end{equation}
	It is obvious that $A$ and $B$ are not $\hat{\pi}$-symmetric but 
	\begin{equation}\label{351}
	\tilde{A}=\left( A+iB\right)^{-1}A= \frac {1}{3} \left(  
	{\begin{array}{rrrrrr}
		2  & 0 & -1 & 0 & -1 & 0 \\ 
		0 & 0 & 0 & 0 & 0 & 0 \\
		-1 & 0 & 2 & 0 &-1 & 0 \\ 
		0 & 0 & 0 & 0 & 0 & 0 \\
		-1 & 0 & -1& 0 & 2 & 0 \\
		0 & 0 & 0 & 0 & 0 & 0
		\end{array}}
	\right), \ \ \
	\end{equation}
	\begin{equation}\label{352}
	\tilde{B}=\left( A+iB\right)^{-1}B= \frac {1}{3i}  \left( 
	{\begin{array}{rrrrrr}
		1 & 0 & 1 & 0 & 1 & 0 \\ 
		0 & 3 & 0 & 0 & 0 & 0 \\
		1 & 0 & 1 & 0 & 1 & 0 \\ 
		0 & 0 & 0 & 3 & 0 & 0 \\
		1 & 0 & 1 & 0 & 1 & 0 \\ 
		0 & 0 & 0 & 0 & 0 & 3
		\end{array}}
	\right)
	\end{equation}
	are $\hat{\pi}$-symmetric and give the equivalent vertex conditions (see Section~\ref{sec23}). 
	
	Now, we choose the regular representation $ \rho:G\longrightarrow GL_{2}\left(\mathbb{C}\right) $ of the group $C_2$ to construct a quotient quantum graph $\Gamma_\rho$. The regular representation of $G=C_2$ can be expressed 
	$$
	\rho (I) := I_2 ,  \ \ 
	\rho(R) :=  \left(
	\begin{array}{rr}
	0 & 1 \\
	1 & 0
	\end{array}
	\right),
	$$
	by choosing the ordered basis $ G=\{I, R \} $ for the space $ \mathbb{C}G $. 
	Now we find a basis of the space $ \Hom_{G}\left(\mathcal{V}_{\rho},\mathcal{V}_{\pi}\right)=\Hom_{G}\left(\mathbb{C}^{2},\mathbb{C}^{3}\right) $. Assume 
	$$ \phi=\begin{pmatrix}
	a & b \\c & d \\e & f  	
	\end{pmatrix}\in \Hom_{G}\left(\mathbb{C}^{2},\mathbb{C}^{3}\right).  $$ 
	
	It follows from the intertwining property (\ref{int}) that
	\begin{equation*}
	\pi\left( R \right) \phi=\phi\rho\left( R \right) \Rightarrow \begin{pmatrix}
	a & b \\e & f \\c & d  	
	\end{pmatrix}=\begin{pmatrix}
	b & a \\d & c \\f & e  	
	\end{pmatrix} \Rightarrow a=b, d=e, c=f,
	\end{equation*}
	which implies 
	\begin{equation}\label{345}
	\left\lbrace \begin{pmatrix}
	1 & 1 \\0 & 0 \\0 & 0  	
	\end{pmatrix},  \begin{pmatrix}
	0 & 0 \\1 & 0 \\0 & 1  \end{pmatrix}, \begin{pmatrix}
	0 & 0 \\0 & 1 \\1 & 0   		
	\end{pmatrix} \right\rbrace, 
	\end{equation} 
	is a basis for $ \Hom_{G}\left(\mathbb{C}^{2},\mathbb{C}^{3}\right) $. Applying vectorization map to these basis elements and then normalizing we get $ \Theta $ in \eqref{207}
	\begin{equation*}
	\Theta= \dfrac{1}{\sqrt{2}}\begin{pmatrix}
	1 & 0 & 0 \\0 & 1 & 0 \\0 & 0 & 1 \\ 1 & 0 & 0 \\0 & 0 & 1 \\0 & 1 & 0  	
	\end{pmatrix},
	\end{equation*}
	whose columns form an orthonormal basis for $ K_{G}\left( \rho, \pi \right) $. Note that   
	\begin{equation*}
	\Theta= \dfrac{1}{\sqrt{|G|}}\begin{pmatrix}
	\pi \left( I\right) \\\pi \left( R\right)  	
	\end{pmatrix},
	\end{equation*}
	which gives us the fundamental idea for our solution that $ \Theta $ in \eqref{207} can always be represented 
	\begin{equation*}
	\Theta=\dfrac{1}{\sqrt{|G|}} \begin{pmatrix}
	\pi \left( g_{1}\right) \\\pi \left( g_{2}\right)\\\vdots\\
	\pi \left( g_{|G|}\right)  	
	\end{pmatrix}.
	\end{equation*}
	
	Note that the order of the blocks is not unique since one can choose another ordering for the basis elements of $ K_{G}\left( \rho, \pi \right) $, but from the definitions of the vertex conditions in (\ref{207}) it is easily seen that the resulting matrices $ A\rho $ and $ B\rho $ do not depend on the order of the blocks. 
	
	Coming back to our example, we choose the fundamental domain $ D=\{e_{1},e_{2}\} $ and we have 
	\begin{equation*}
	O_{1}=\{e_{1}\},\ O_{2}=\{e_{2},e_{3}\},\ G_{e_{1}}=G,\ G_{e_{2}}=\{I\}.
	\end{equation*}
	Obviously, $ d_{2}=2 $ and from (\ref{211}) we get
	\begin{equation*}
	d_{1}=\dim \bigcap _{g \in G} \ker \left[ I_{2}-\rho\left( g \right)^{T} \right]=1.
	\end{equation*}
	Indeed, assume $ f=\left( f_{1}, f_{2}\right)^{T} 
	$ satisfies the equalities
	\begin{equation*}
	I_{2}f=\rho\left( I \right)^{T}f, \quad I_{2}f=\rho\left( R \right)^{T}f.
	\end{equation*}
	It follows $ \left( f_{1}, f_{2}\right)^{T}=\left( f_{2}, f_{1}\right)^{T} $ and hence $ \left\lbrace \left( 1,1\right)^{T} \right\rbrace $ is a basis for 
	$$ \bigcap _{g \in G} \ker \left[ I_{2}-\rho\left( g \right)^{T} \right] $$.
	Therefore, we can construct the quotient graph $\Gamma_\rho$ which has 
	$$ 
	d=\dim \Hom_{G}\left(\mathbb{C}^{2},\mathbb{C}^{3}\right)=d_{1}+d_{2}=3
	$$ 
	edges which consist of $d_{2}=2  $ copies of $ e_{2} $ and $d_{1}=1  $ copy of $ e_{1} $. One can easily compute that the vertex conditions are given by $ A_\rho = \tilde{A} $ and $ B_\rho = \tilde{B} $ (see (\ref{207})) for the quotient quantum graph $\Gamma_\rho$. Therefore, we obtain a quotient graph $\Gamma_\rho$ which is identical to the original quantum graph $\Gamma$ if we choose $ G=\{I, R \} $ as a basis of $ \mathbb{C}G $.  
\end{example}	
\section{Main results}	
In this section, we solve the open problem stated in the last section affirmatively in the light of the observations made for Example~\ref{example1}. More explicitly, we prove that if one constructs the quotient graph $ \Gamma/\mathbb{C}G $ by choosing $ G $ as a basis for $ \mathbb{C}G $, one obtains the quotient graph $ \Gamma/\mathbb{C}G $ which is identical to $ \Gamma $. In other words, $ \Gamma $ is a $ \Gamma/\mathbb{C}G $ graph. We begin by proving a more general result that is $ \Gamma $ is a $ \Gamma/\rho $ graph for an arbitrary permutation representation $ \rho $ of $ G $ with degree $ |G| $. We also show by a counterexample that this is not necessarily true for a permutation representation of $ G $ with degree greater than $ |G| $. 
\begin{theorem} \label{b}
	Let $ \Gamma $ be a $ \pi $-symmetric quantum graph where $ \pi:G \longrightarrow GL_{|E|}\left( \mathbb{C} \right)$ is a permutation representation of the finite group $ G $ acting on $ \Gamma $ with vertex conditions given by (\ref{204}). Let $ \rho$ be an arbitrary permutation representation of $ G $ with degree $ |G| $. Let us choose the standard basis of $ \mathbb{C}^{|G|} $ as a basis for $ \rho$ and any fundamental domain for the action of $ G $. Then, the quotient graph $ \Gamma_{\rho} $ is identical to $ \Gamma $ for these choice. 
\end{theorem}
\begin{proof}
	Note that the matrices $ A $ and $ B $ in (\ref{204}) are assumed to be $ \hat{\pi} $-symmetric (see Section~\ref{sec23}). Let us choose a fundamental domain $ D=\{e_{1},e_{2},\ldots, e_{|D|}\} $ and let $ \rho$ be an arbitrary permutation representation of degree $ r:=|G| $. Since $ \rho$ is a permutation representation, each $ g \in G $ permutes the standard basis elements $ e_{i} $ of $  \mathbb{C}^{r}$ and naturally $ G $ acts on the set of indices $ \left\lbrace 1,2,...,r\right\rbrace $ such that $ g\cdot e_{i}:=e_{g \cdot i} $. Assume without loss of generality that elements of $ G $ are labeled $ \{g_{1},g_{2},\ldots,g_{r} \}$ such that $ g_{j}.1=j \ (j=1,2,...,r) $. Then it easily follows $ \rho \left( g_{j} \right) e_{1}=e_{g_{j}\cdot 1}=e_{j} $.
	
	We begin by establishing the basis	of $ \Hom_{G}\left(\mathcal{V}_{\rho},\mathcal{V}_{\pi}\right) $ and show that $ \Theta $ in Definition \ref{def} can be represented 
	\begin{equation*}
	\Theta=\dfrac{1}{\sqrt{r}} \begin{pmatrix}
	\pi \left( g_{1}\right) \\\pi \left( g_{2}\right)\\\vdots\\
	\pi \left( g_{r}\right)  	
	\end{pmatrix},
	\end{equation*}   
	if we use the standard basis of $ \mathbb{C}^{|G|} $ as a basis for $ \rho$. Assume 
	$$ f=\left( f_{1} \ f_{2} \ \cdots \ f_{r}\right)\in \Hom_{G}\left(\mathcal{V}_{\rho},\mathcal{V}_{\pi}\right),$$ 
	where $ f_{k}\in \mathbb{C}^{p}\ \left( k=1,2,\ldots,r \right)  $ with $ p:=\dim \mathcal{V}_{\pi} =|E|$ and $ \ r=\dim \mathcal{V}_{\rho}=|G|$. Then, $ r $-tuple of vectors $ f_{k}\in \mathbb{C}^{p} $, $ k=1,2,\ldots,r $ transforms under $ \pi $ according to $ \rho $ (see Section~\ref{sec22})
	\begin{equation*}
	\pi\left( g\right)f_{k}=\sum_{i=1}^{r}\left[\rho\left( g\right) \right]_{ik}f_{i}  \\
	=f_{g.k}, \quad \forall g\in G, \quad k=1,2,\ldots,r  
	\end{equation*} 
	where $ g\cdot k $ corresponds to $ \rho\left( g\right) e_{k}=g \cdot e_{k}=e_{g\cdot k} $. Basically, $ \pi\left( g\right) $ permutes the columns of $ f $. From the assumption $ g_{j}\cdot 1=j \ (j=1,2,...,r) $ we can write 
	\begin{align*}
	f &= \left( f_{1} \ f_{2} \ \cdots \ f_{r}\right) \\
	&=\left( \pi( g_{1})f_{1} \quad \pi( g_{2})f_{1} \quad \cdots \quad \pi( g_{r})f_{1} \right)\\
	&=f_{11}\left( \pi\left( g_{1}\right)_{1} \quad \pi\left( g_{2}\right)_{1} \quad \cdots \quad \pi\left( g_{r}\right)_{1} \right) \\ 
	& +f_{12}\left( \pi\left( g_{1}\right)_{2} \quad \pi\left( g_{2}\right)_{2} \quad \cdots \quad \pi\left( g_{r}\right)_{2} \right) \\
	& + \cdots+ f_{1p} \left( \pi\left( g_{1}\right)_{p} \quad \pi\left( g_{2}\right)_{p} \quad \cdots  \quad \pi\left( g_{r}\right)_{p} \right),
	\end{align*}
	where $ f_{1}=\left( f_{11},f_{12},\cdots,f_{1p} \right)^{T} $ and $ \pi\left( g\right)_{i}  $ denotes the $ i $-th column of $ \pi\left( g\right) $. Hence 
	\begin{align*}
	\left\lbrace \left( \pi \left( g_{1}\right)_{1}\ \pi \left( g_{2}\right)_{1} \ \cdots\ \pi\left( g_{r}\right)_{1} \right), \left( \pi\left( g_{1}\right)_{2}\ \pi\left( g_{2}\right)_{2}\ \cdots\ \pi\left( g_{r}\right)_{2} \right) \right. \\ \left. , \ldots , \left( \pi \left( g_{1}\right)_{p}\ \pi\left( g_{2}\right)_{p}\ \cdots\ \pi\left( g_{r}\right)_{p} \right) \right\rbrace 
	\end{align*}
	is a basis for $ \Hom_{G} \left(\mathcal{V}_{\rho},\mathcal{V}_{\pi}\right) $ and $ d=\dim \Hom_{G}\left(\mathcal{V}_{\rho},\mathcal{V}_{\pi}\right)=p. $
	By applying the vectorization map to these basis elements we obtain the matrix 
	\begin{equation*}
	\Theta= \dfrac{1}{\sqrt{r}}\begin{pmatrix}
	\pi\left( g_{1}\right)_{1} \pi\left( g_{1}\right)_{2}\cdots\pi\left( g_{1}\right)_{p} \\\pi\left( g_{2}\right)_{1} \pi\left( g_{2}\right)_{2}\cdots\pi\left( g_{2}\right)_{p}\\\vdots\\
	\pi\left( g_{r}\right)_{1} \pi\left( g_{r}\right)_{2}\cdots\pi\left( g_{r}\right)_{p}  	
	\end{pmatrix},
	\end{equation*} 
	whose columns form an orthonormal basis for $ K_{G}\left( \rho, \pi \right) $. We can rewrite $ \Theta $ as
	\begin{equation}\label{455}
	\Theta=\dfrac{1}{\sqrt{r}} \begin{pmatrix}
	\pi \left( g_{1}\right) \\\pi \left( g_{2}\right)\\\vdots\\
	\pi \left( g_{r}\right)  	
	\end{pmatrix}
	\end{equation} 
	and also
	\begin{equation*}
	\hat{\Theta}=\Theta\otimes I_{2}=\dfrac{1}{\sqrt{r}} \begin{pmatrix}
	\pi \left( g_{1}\right)\otimes I_{2} \\\pi \left( g_{2}\right) \otimes I_{2}\\\vdots\\
	\pi \left( g_{r}\right) \otimes I_{2}  	
	\end{pmatrix}=\dfrac{1}{\sqrt{r}}\begin{pmatrix}
	\hat{\pi} \left( g_{1}\right) \\\hat{\pi} \left( g_{2}\right)\\\vdots\\
	\hat{\pi} \left( g_{r}\right)  	
	\end{pmatrix}. 
	\end{equation*}
	By the definition of the quotient quantum graph $ \Gamma_{\rho} $, the vertex conditions are given by the equations (\ref{207}), namely:
	\begin{align*}
	A_{\rho} &= \hat{\Theta}^{\ast} \left[ I_{r}\otimes A \right]\hat{\Theta} \\ 
	&= \frac{1}{r}\begin{pmatrix}
	\hat{\pi}^{\ast} \left( g_{1}\right) \cdots \
	\hat{\pi}^{\ast} \left( g_{r}\right)  	
	\end{pmatrix}diag \left( A,A,\ldots,A \right)\begin{pmatrix}
	\hat{\pi} \left( g_{1}\right) \\\hat{\pi} \left( g_{2}\right)\\\vdots\\
	\hat{\pi} \left( g_{r}\right)  	
	\end{pmatrix} \\ 
	&= \frac{1}{r}\left[ 
	\hat{\pi}^{\ast} \left( g_{1}\right)A\hat{\pi} \left( g_{1}\right)+ \cdots +
	\hat{\pi}^{\ast} \left( g_{r}\right)A\hat{\pi} \left( g_{r}\right) \right] \\
	&= \dfrac{1}{r}\left[ A+\cdots+A \right]\\
	&= A,    
	\end{align*}
	since $ A $ is $ \hat{\pi} $-symmetric. Similarly, we find $ B_{\rho}=B $.
	
	The last step of the proof is to show that the number $ d_i $ of the copies of the edge $ e_i $ in the quotient equals $ |O_{i}| $. Recall that from (\ref{211})
	\begin{equation}\label{408}
	d_{i}=\dim \bigcap _{g \in G_{e_{i}}} \ker \left[ I_{r}-\rho\left( g \right)^{T} \right]. 
	\end{equation}
	Let $ e \in E $. If the stabilizer $ G_{e} $ is trivial, then by (\ref{408}) it easily follows $ d_i=r $. Now, let us assume $ G_{e}\neq \{id\} $ and $ G_{e}=\{g_{1}=id, g_{2},\ldots,g_{k}\} $ without loss of generality where $ id $ denotes the identity element of $ G $. Now let 
	\begin{equation*}
	f=\left( f_{1},f_{2},\ldots, f_{r}\right)^{T}   	
	\in \bigcap _{g \in G_{e}} \ker \left[ I_{r}-\rho\left( g \right)^{T} \right]. 
	\end{equation*} 
	Then
	\begin{equation*}
	f=\rho\left( g_{i}\right)^{T}f,\quad  i=1,2,\ldots,k.
	\end{equation*}  
	Since $ \rho\left( g_{i}\right)^{T} $ is a permutation matrix it permutes the components of $ f $ in a way
	\begin{equation*}
	\rho\left( g_{i}\right)^{T}f=\left( 
	f_{g_{i}.1},f_{g_{i}.2},\ldots,f_{g_{i}.r}  	
	\right)^{T} .
	\end{equation*}
	This implies $$ f_{j}=f_{g_{i}.j} , \quad i=1,2,\ldots,k, \quad j=1,2,\ldots,r  .$$ Now, this suggests that $$ d_{i}= \frac{r}{k}=\left[ G :G_{e} \right]=|O_{e}| $$ as desired. As a result, $ \Gamma_{\rho} $ is identical to $ \Gamma $.  
\end{proof}
\begin{remark}
	The results in Theorem~\ref{b} do not hold for a permutation representation of degree greater than $ \lvert G\lvert. $ We demonstrate this with an example. Consider the quantum graph in Example~\ref{example1}. Let us define the permutation representation
	\begin{equation*}
	\rho \left( R \right) =\begin{pmatrix}
	0 & 1 & 0\\1 & 0 & 0 \\ 0 & 0 & 1   	
	\end{pmatrix}
	\end{equation*} 
	with degree 3. From the intertwining property (\ref{int}) we obtain a basis 
	\begin{equation*}
	\left\lbrace \begin{pmatrix}
	1 & 1 & 0  \\ 0 & 0 & 0 \\0 & 0 & 0  	
	\end{pmatrix},  \begin{pmatrix}
	0 & 0 & 0 \\0 & 1 & 0 \\1 & 0 & 0  \end{pmatrix}, \begin{pmatrix}
	0 & 0 & 0 \\1 & 0 & 0 \\0 & 1 & 0   		
	\end{pmatrix}, \begin{pmatrix}
	0 & 0 & 0 \\0 & 0 & 1 \\0 & 0 & 1   		
	\end{pmatrix} \right\rbrace, 
	\end{equation*}
	for $ K_{G}\left( \rho, \pi \right) $ which implies $ d=4 $ and this contradicts the results of the theorem.
\end{remark} 
As a result of Theorem \ref{b} we can prove that $ \Gamma $ is a $ \Gamma/\mathbb{C}G $ graph and hence answer the open problem stated in the last section affirmatively.
\begin{corollary} \label{ana}
	Let $ \Gamma $ be a $ \pi $-symmetric quantum graph where $ \pi:G \longrightarrow GL_{|E|}\left( \mathbb{C} \right)$ is a permutation representation of the finite group $ G $ acting on $ \Gamma $ with vertex conditions given by (\ref{204}). Let $ \rho$ denote the regular representation $ \rho_{G} $ of $ G $. Let us choose the ordered basis $ G= \{g_{1},g_{2},\ldots,g_{r} \}$ as a basis for $ \mathcal{V}_{\rho}=\mathbb{C}G $ and any fundamental domain for the action of $ G $. Then, $ \Gamma_{\rho} $ is identical to $ \Gamma $ for these choice.   
\end{corollary}
\begin{proof}
	Note that the matrices $ A $ and $ B $ in (\ref{204}) are assumed to be $ \hat{\pi} $-symmetric (see Section~\ref{sec23}). Let us choose a fundamental domain $ D=\{e_{1},e_{2},\ldots, e_{|D|}\} $ and the ordered basis $ G= \{g_{1},g_{2},\ldots,g_{r} \}$ as a basis for $ \mathbb{C}G $. Since the regular representation of $ G $ is a permutation representation of degree $ |G| $ it follows from Theorem \ref{b} that $ \Gamma_{\rho} $ is identical to $ \Gamma $ for these choice and the proof is done. However, we want to mention that we can show similarly as in Theorem \ref{b} that $ \Theta $ in Definition \ref{def} can be represented 
	\begin{equation}\label{888}
	\Theta=\dfrac{1}{\sqrt{|G|}} \begin{pmatrix}
	\pi \left( g_{1}g_{1}^{-1}\right) \\\pi \left( g_{2}g_{1}^{-1}\right)\\\vdots\\
	\pi \left( g_{r}g_{1}^{-1}\right)  	
	\end{pmatrix},
	\end{equation}  
	if we use the ordered basis $ \{g_{1},g_{2},\ldots,g_{r} \}$ of $ \mathbb{C}G $. This representation of $ \Theta $ will be used in examples in the next section. We also would like to point out that the number $ d_i $ of the copies of the edge $ e_i $ in the quotient equals $ |O_{i}| $ which can be proved similarly as in Theorem \ref{b}. This property will be also used in examples in the next section.
\end{proof}

\section{Additional examples}
In this section, we provide some examples to demonstrate how our main results work.

\begin{example}
	We reconsider Example~\ref{example1} presented in Section~\ref{sec3}. This time, we would like to take the same basis in (\ref{345}) with a different ordering
	\begin{equation*}
	\left\lbrace \begin{pmatrix}
	1 & 1 \\0 & 0 \\0 & 0  	
	\end{pmatrix}, \begin{pmatrix}
	0 & 0 \\0 & 1 \\1 & 0   		
	\end{pmatrix}, \begin{pmatrix}
	0 & 0 \\1 & 0 \\0 & 1  \end{pmatrix} \right\rbrace.  
	\end{equation*}
	Applying vectorization to these basis elements and then normalizing we get $ \Theta  $ in \eqref{207}
	\begin{equation*}
	\Theta= \dfrac{1}{\sqrt{2}}\begin{pmatrix}
	1 & 0 & 0 \\0 & 0 & 1 \\0 & 1 & 0 \\ 1 & 0 & 0 \\0 & 1 & 0 \\0 & 0 & 1 	
	\end{pmatrix},
	\end{equation*}
	whose columns form an orthonormal basis for $ K_{G}\left( \rho, \pi \right) $. Note that   
	\begin{equation*}
	\Theta= \dfrac{1}{\sqrt{|G|}}\begin{pmatrix}
	\pi \left( R\right) \\ \pi \left( I\right)
	\end{pmatrix}.
	\end{equation*}
	Note also that the order of the blocks of $ \Theta $ is now interchanged but the resulting matrices $ A_\rho $ and $ B_\rho $ in \eqref{207} remain the same.
\end{example} 

\begin{example}
	We consider a star graph with 3 edges having the same length and equipped with Neumann boundary conditions (see Figure 2).
	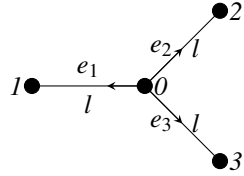
\begin{figure} [h]
		\centering	
		\begin{tikzpicture}
		\coordinate [label=right:0] (0) at (0,0);
		\coordinate [label=left:1] (1) at (-1.5,0);
		\coordinate [label=right:2] (2) at (1,1);
		\coordinate [label=right:3] (3) at (1,-1);
		\draw (0)--(1);
		\draw (0)--(2);
		\draw (0)--(3);
		
		\draw [-stealth] (0,0)--(-0.5,0);
		\draw [-stealth] (0,0)--(0.5,0.5);
		\draw [-stealth] (0,0)--(0.5,-0.5);
		
		\draw[fill] (0,0) circle [radius=0.1];
		\draw[fill] (-1.5,0) circle [radius=0.1];
		\draw[fill] (1,1) circle [radius=0.1];
		\draw[fill] (1,-1) circle [radius=0.1];
		
		\coordinate[label=below:$l$](01) at ($ (0)!.5!(1) $);
		\coordinate[label=right:$l$] (03) at ($ (0)!.5!(3) $);
		\coordinate[label=right:$l$](02) at ($ (0)!.5!(2) $);
		\coordinate[label=above:$e_1$](01) at ($ (0)!.5!(1) $);
		\coordinate[label=left:$e_2$](02) at ($ (0)!.5!(2) $);
		\coordinate[label=left:$e_3$](03) at ($ (0)!.5!(3) $);
		\end{tikzpicture}
		\caption{A star graph with 3 edges having the same length. }
	\end{figure}
	Let $ R $ denote the counter-clockwise rotation by 120$ ^{\circ} $ around the central vertex and take the cyclic group $ G=\left\lbrace I,R,R^{2} \right\rbrace  $ generated by $ R $ . Clearly $ G $ acts as a symmetry group on $ \Gamma $. Let us define a representation of this group by
	\begin{equation*}
	\pi\left( R \right) =\begin{pmatrix}
	0 & 1 & 0 \\0 & 0 & 1 \\1 & 0 & 0  	
	\end{pmatrix}.
	\end{equation*} 
	The vertex conditions can be given by the matrices $A$ and $B$ in \eqref{350}. The matrices $A$ and $B$ are not $\hat{\pi}$-symmetric but $ \tilde{A} $ and $ \tilde{B} $ in Equations (\ref{351}) and (\ref{352}) are $\hat{\pi}$-symmetric (see Example~\ref{example1}).	Now, we consider the regular representation $ \rho:G\longrightarrow GL_{3}\left(\mathbb{C}\right) $ of the group $G$ to construct a quotient quantum graph $\Gamma_\rho$. The regular representation of $G$ can be expressed
	\begin{equation*}
	\rho \left( R \right) =\begin{pmatrix}
	0 & 0 & 1\\1 & 0 & 0 \\ 0 & 1 & 0   	
	\end{pmatrix},
	\end{equation*} 
	by choosing the ordered basis for the space $ \mathbb{C}G $ as $$ G=\{g_{1}=I, g_{2}=R, g_{3}=R^{2} \}. $$ It is easy to show that 
	\begin{equation*}
	\left\lbrace \begin{pmatrix}
	1 & 0 & 0  \\ 0 & 0 & 1 \\0 & 1 & 0  	
	\end{pmatrix},  \begin{pmatrix}
	0 & 1 & 0 \\1 & 0 & 0 \\0 & 0 & 1  \end{pmatrix}, \begin{pmatrix}
	0 & 0 & 1 \\0 & 1 & 0 \\1 & 0 & 0   		
	\end{pmatrix} \right\rbrace, 
	\end{equation*} 
	is a basis for $ \Hom_{G}\left(\mathcal{V}_{\rho},\mathcal{V}_{\pi}\right) $. Hence applying vectorization to these basis elements and then normalizing we get 
	\begin{equation*}
	\Theta= \dfrac{1}{\sqrt{3}}\begin{pmatrix}
	1 & 0 & 0 \\0 & 1 & 0 \\0 & 0 & 1 \\0 & 1 & 0\\0 & 0 & 1\\ 1 & 0 & 0 \\0 & 0 & 1 \\ 1 & 0 & 0\\0 & 1 & 0  \end{pmatrix},
	\end{equation*}
	whose columns form an orthonormal basis for $ K_{G}\left( \rho, \pi \right) $. We have   
	\begin{equation*}
	\Theta= \dfrac{1}{\sqrt{|G|}}\begin{pmatrix}
	\pi \left( g_{1}\right) \\\pi \left(g_{2}\right) \\\pi \left(g_{3}\right)  	
	\end{pmatrix}.
	\end{equation*}
	Let us choose the fundamental domain $ D=\{e_{1}\} $. Then, we have $ O_{1}=\{e_{1},e_{2},e_{3}\}.$ Since the stabilizer $ G_{e_{1}} $ is trivial, it follows $ d_{1}=3. $ 
	Therefore, we can construct the quotient graph $\Gamma_\rho$ which has 3 copies of the edge $ e_{1}.  $ One can easily see that $ A_\rho = \tilde{A} $ and $ B_\rho = \tilde{B} $ (see (\ref{207})). The resulting quotient graph $\Gamma_\rho$ is identical to $\Gamma$.
	
	Now, consider a permutation representation of $ G $ with degree 3 given by 
	\begin{equation*}
	\sigma \left( R \right) =\begin{pmatrix}
	0 & 1 & 0 \\0 & 0 & 1 \\1 & 0 & 0  	
	\end{pmatrix}.
	\end{equation*} 
	From the intertwining property (\ref{int}) we obtain $\Theta  $ in \eqref{207} by an easy calculation 
	\begin{equation*}
	\Theta= \dfrac{1}{\sqrt{3}}\begin{pmatrix}
	1 & 0 & 0 \\0 & 1 & 0 \\0 & 0 & 1 \\0 & 0 & 1\\1 & 0 & 0\\ 0 & 1 & 0 \\0 & 1 & 0 \\ 0 & 0 & 1\\1 & 0 & 0  \end{pmatrix},
	\end{equation*}
	which can be written 
	\begin{equation*}
	\Theta= \dfrac{1}{\sqrt{|G|}}\begin{pmatrix}
	\pi \left( I\right) \\\pi \left( R^{2} \right) \\\pi \left(R \right)  	
	\end{pmatrix}.
	\end{equation*}
	Now let us to find out this with the help of our results in Theorem~\ref{b}. We have 
	\begin{equation}\nonumber
	\sigma \left( I \right)e_{1}=e_{1}, \ \sigma \left( R \right)e_{1}=e_{3}, \ \sigma \left( R^{2} \right)e_{1}=e_{2},
	\end{equation}
	which implies $ g_{1}=I, \ g_{2}=R^{2}, \ g_{3}=R $ in (\ref{455}) and 
	\begin{equation}\nonumber
	\Theta= \dfrac{1}{\sqrt{3}}\begin{pmatrix}
	\pi \left( g_{1}\right) \\\pi \left(g_{2}\right) \\\pi \left(g_{3} \right)  	
	\end{pmatrix}.
	\end{equation}
	Again, we have $ A_\sigma = \tilde{A} $ and $ B_\sigma = \tilde{B} $ and we conclude $\Gamma_\sigma $ is identical to $\Gamma$. 
\end{example} 

\begin{example}
	We reconsider the same quantum graph given in Figure 2 but the symmetry group is now the symmetric group 
	\begin{equation*}
	G:=S_3=\{ g_{1}=(1), g_{2}=\sigma, g_{3}=\sigma^2, g_{4}=\tau, g_{5}=\sigma \tau, g_{6}=\sigma^2 \tau \},
	\end{equation*}
	where $\sigma=(123)$ and $\tau=(12)$. Clearly, the graph is $ \pi -$symmetric where $ \pi $ is given 
	$$
	\pi(I)=I_3, \    
	\pi(\sigma)= \left( \begin{matrix} 
	0 & 0 & 1 \\
	1 & 0 & 0 \\
	0 & 1 & 0 
	\end{matrix}
	\right), \ 
	\pi(\sigma^2)= \left( \begin{matrix} 
	0 & 1 & 0 \\
	0 & 0 & 1 \\
	1 & 0 & 0 
	\end{matrix}
	\right),$$
	
	$$
	\pi(\tau)= \left( \begin{matrix} 
	0 & 1 & 0 \\
	1 & 0 & 0 \\
	0 & 0 & 1 
	\end{matrix}
	\right), 
	\pi(\sigma \tau)= \left( \begin{matrix} 
	0 & 0 & 1 \\
	0 & 1 & 0 \\
	1 & 0 & 0 
	\end{matrix}
	\right),
	\pi(\sigma^2 \tau)= \left( \begin{matrix} 
	1 & 0 & 0 \\
	0 & 0 & 1 \\
	0 & 1 & 0 
	\end{matrix}
	\right). 
	$$ 
	Now, we take the regular representation $ \rho $ of $S_3$ which is given by
	\begin{equation*}
	\rho(\sigma)= \left( \begin{matrix} 
	0 & 0 & 1 & 0 & 0 & 0 \\
	1 & 0 & 0 & 0 & 0 & 0 \\
	0 & 1 & 0 & 0 & 0 & 0 \\
	0 & 0 & 0 & 0 & 1 & 0 \\
	0 & 0 & 0 & 0 & 0 & 1 \\
	0 & 0 & 0 & 1 & 0 & 0
	\end{matrix}
	\right), \ \rho(\tau)= \left( \begin{matrix} 
	0 & 0 & 0 & 0 & 0 & 1 \\
	0 & 0 & 0 & 1 & 0 & 0 \\
	0 & 0 & 0 & 0 & 1 & 0 \\
	0 & 1 & 0 & 0 & 0 & 0 \\
	0 & 0 & 1 & 0 & 0 & 0 \\
	1 & 0 & 0 & 0 & 0 & 0
	\end{matrix}
	\right),
	\end{equation*} 
	where the ordered basis for $ \mathbb{C}G $ is taken $ G=\{g_{1}, g_{2}, g_{3}, g_{4}, g_{5}, g_{6} \} $. We use Maple to find an orthonormal basis for $ K_{G}\left( \rho, \pi \right) $ and apply vectorization map to these basis elements. We obtain the matrix $ \Theta $ in \eqref{207}
	\begin{equation*}
	\Theta= \dfrac{1}{\sqrt{6}}\begin{pmatrix}
	\pi \left( g_{6}\right) \\\pi \left(g_{4}\right) \\\pi \left(g_{5}\right) \\ \pi \left(g_{1}\right) \\ \pi \left(g_{3}\right) \\ \pi \left(g_{2}\right)	
	\end{pmatrix}.
	\end{equation*}
	Using \eqref{888} we obtain 
	\begin{equation*}
	\Theta= \dfrac{1}{\sqrt{6}}\begin{pmatrix}
	\pi \left( g_{1}\right) \\\pi \left(g_{2}\right) \\\pi \left(g_{3}\right) \\\pi \left(g_{4}\right)  \\ \pi \left(g_{5}\right) \\ \pi \left(g_{6}\right)	
	\end{pmatrix}.
	\end{equation*}
	Note that the difference between two matrices is just the order of the blocks which depends on the ordering of the basis elements of $ K_{G}\left( \rho, \pi \right) $. Nonetheless, the matrices $ A_\rho $ and $ B_\rho $ in \eqref{207} are the same and explicitly $A_\rho= \tilde{A}, \quad B_\rho = \tilde{B} $. Let us choose the fundamental domain $ D=\{e_{1}\} $. Then, we have $ O_{1}=\{e_{1},e_{2},e_{3}\}$ . The stabilizer of the edge $ e_{1}$ is $ G_{e_{1}}=\left\lbrace g_{1},g_{6} \right\rbrace$ and we have $ d_{1}=\dfrac{6}{2}=3 $ (see Corollary~\ref{ana}). The quotient graph $\Gamma_\rho$ has 3 copies of the edge $ e_{1}$ and is identical to $\Gamma$ as desired.
\end{example}
\begin{example} \label{ex54}
	Our last example might be the equilateral tetrahedron graph $ \Gamma $ (see Figure 3) with Neumann boundary conditions. 
	\begin{figure} [h]
		\centering	
		\begin{tikzpicture}
		\coordinate [label=left:1] (1) at (0,0);
		\coordinate [label=right:4] (4) at (4,0);
		\coordinate [label=above:2] (2) at (2,3.4641);
		\coordinate [label=below:3] (3) at (2,1.1547);
		\draw (1)--(2)--(3)--cycle;
		\draw (1)--(4)--(3);
		\draw (4)--(2);
		\draw[fill] (0,0) circle [radius=0.07];
		\draw[fill] (4,0) circle [radius=0.07];
		\draw[fill] (2,3.4641) circle [radius=0.07];
		\draw[fill] (2,1.15471) circle [radius=0.07];
		\coordinate[label=below:$l$](12) at ($ (1)!.5!(2) $);
		\coordinate[label=left:$l$] (13) at ($ (1)!.5!(3) $);
		\coordinate[label=right:$l$](23) at ($ (2)!.5!(3) $);
		\coordinate[label=below:$l$](14) at ($ (1)!.5!(4) $);
		\coordinate[label=below:$l$](42) at ($ (4)!.5!(2) $);
		\coordinate[label=right:$l$](34) at ($ (3)!.5!(4) $);
		\end{tikzpicture}
		\caption{ An equilateral tetrahedron graph.  }
	\end{figure}
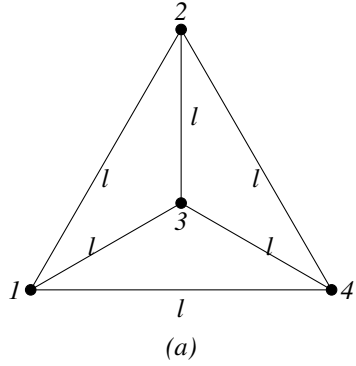
	Its symmetry group is $S_4$ which acts non-freely on the edges. For the vertex conditions, we can write the matrices $A$ and $B$ in \eqref{q1} such that 
	\begin{align*}
	&A_{1,1}=A_{2,3}=A_{3,17}=A_{4,19}=A_{5,11}=A_{6,13}=A_{7,7}=1,\\
	&A_{8,9}=A_{9,6}=A_{10,4}=A_{11,10}=A_{12,2}=A_{13,16}=A_{14,22}=1,\\
	&A_{1,3}=A_{2,5}=A_{3,19}=A_{4,21}=A_{5,13}=A_{6,15}=A_{7,9}=-1, \\ 
	&A_{8,23}=A_{9,8}=A_{10,14}=A_{11,12}=A_{12,20}=A_{13,18}=A_{14,24}=-1,
	\end{align*}
	and
	\begin{align*}
	&B_{15,1}=B_{15,3}=B_{15,5}=B_{16,17}=B_{16,19}=B_{16,21}=B_{17,11}=B_{17,13}=1,\\
	&B_{17,15}=B_{18,7}=B_{18,9}=B_{18,23}=B_{19,6}=B_{19,8}=B_{20,4}=B_{20,14}=1,\\
	&B_{21,10}=B_{21,12}=B_{22,2}=B_{22,20}=B_{23,16}=B_{23,18}=B_{24,22}=B_{24,24}=1,
	\end{align*} 
	and the other entries of $A$ and $B$ are zero.	
	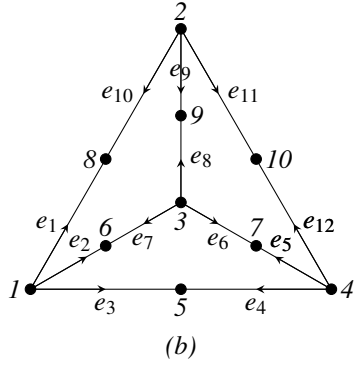
\begin{figure} [h]
		\centering	
		\begin{tikzpicture}
		\coordinate [label=left:1] (n1) at (6,0);
		\coordinate [label=right:4] (n4) at (10,0);
		\coordinate [label=above:2] (n2) at (8,3.4641);
		\coordinate [label=below:3] (n3) at (8,1.1547);
		\draw (n1)--(n2)--(n3)--cycle;
		\draw (n1)--(n4)--(n3);
		\draw (n4)--(n2);
		
		\draw [-stealth] (6,0)--(6.5,0.8660);
		\draw [-stealth] (6,0)--(6.75,0.4330);
		\draw [-stealth] (6,0)--(7,0);
		
		\draw [-stealth] (8,3.4641)--(7.5,2.5981);
		\draw [-stealth] (8,3.4641)--(8,2.5981);
		\draw [-stealth] (8,3.4641)--(8.5,2.5981);
		
		\draw [-stealth] (8,1.1547)--(8,1.7321);
		\draw [-stealth] (8,1.1547)--(7.5,0.8660);
		\draw [-stealth] (8,1.1547)--(8.5,0.8660);
		
		\draw [-stealth] (10,0)--(9.5,0.866025);
		\draw [-stealth] (10,0)--(9.25,0.4330);
		\draw [-stealth] (10,0)--(9,0);
		\coordinate[label=left:$e_1$](12) at ($ (n1)!.25!(n2) $);
		\coordinate[label=above:$e_2$](13) at ($ (n1)!.33!(n3) $);
		\coordinate[label=below:$e_3$](14) at ($ (n1)!.25!(n4) $);
		\coordinate[label=right:$e_{12}$](42) at ($ (n4)!.25!(n2) $);
		\coordinate[label=above:$e_5$](43) at ($ (n4)!.33!(n3) $);
		\coordinate[label=below:$e_4$](14) at ($ (n4)!.25!(n1) $);
		\coordinate[label=left:$e_{10}$](12) at ($ (n2)!.25!(n1) $);
		\coordinate[label=above:$e_9$](23) at ($ (n2)!.35!(n3) $);
		\coordinate[label=right:$e_{12}$](42) at ($ (n4)!.25!(n2) $);
		\coordinate[label=above:$e_5$](43) at ($ (n4)!.33!(n3) $);
		\coordinate[label=right:$e_{11}$](24) at ($ (n2)!.25!(n4) $);
		
		\coordinate[label=below:$e_{7}$](31) at ($ (n3)!.25!(n1) $);
		\coordinate[label=below:$e_{6}$](34) at ($ (n3)!.25!(n4) $);
		\coordinate[label=right:$e_{8}$](32) at ($ (n3)!.25!(n2) $);
		
		\coordinate[label=below:5](14) at ($ (n1)!.5!(n4) $);
		\coordinate[label=above:6] (13) at ($ (n1)!.5!(n3) $);
		\coordinate[label=above:7](34) at ($ (n3)!.5!(n4) $);
		\coordinate[label=left:8](12) at ($ (n1)!.5!(n2) $);
		\coordinate[label=right:9](23) at ($ (n2)!.5!(n3) $);
		\coordinate[label=right:10](42) at ($ (n4)!.5!(n2) $);
		\draw[fill] (6,0) circle [radius=0.07];
		\draw[fill] (10,0) circle [radius=0.07];
		\draw[fill] (8,3.4641) circle [radius=0.07];
		\draw[fill] (8,1.15471) circle [radius=0.07];
		\draw[fill] ($ (n1)!.5!(n2) $) circle [radius=0.07];
		\draw[fill] ($ (n2)!.5!(n3) $) circle [radius=0.07];
		\draw[fill] ($ (n3)!.5!(n4) $) circle [radius=0.07];
		\draw[fill] ($ (n1)!.5!(n3) $) circle [radius=0.07];
		\draw[fill] ($ (n1)!.5!(n4) $) circle [radius=0.07];
		\draw[fill] ($ (n2)!.5!(n4) $) circle [radius=0.07];
		\end{tikzpicture}
		\caption{We added dummy vertices and assigned directions arbitrarily.  }	
	\end{figure}			
	Note that we added dummy vertices (see Figure 4) to prevent an edge gets transformed onto its reversal (see Section~\ref{sec23}). We choose the fundamental domain $ D=\{e_{1}\} $. Then, we have $ O_{1}=E=\{e_{1},e_{2},\ldots, e_{12}\}$. The stabilizer of the edge $ e_{1} $ is $ G_{e_{1}}=\left\lbrace (1),(34) \right\rbrace$. Thus, we have $ d_{1}=\dfrac{24}{2}=12 $ copies of the edge $ e_{1} $ to establish the quotient graph (see Corollary~\ref{ana}).\\
	Let $a=(12)$, $b=(23)$ and $c=(34)$. Then, $S_4$ can be written as 
	\begin{align*}
	S_4 &= \left\lbrace aa, c, b, cb, bc, bcb, a, ac, bc, cba, bca, bcba, ab, cab, aba,\right. \\  &  \ \ \left. caba, bcab, bcaba, abcab, abcb, abca, abcba, abcab, abcaba \right\rbrace. 
	\end{align*}
	Letting  
	$$
	\pi (a) :=  \left(  
	{\begin{array}{rrrrrrrrrrrr}
		0 & 0 & 0 & 0 & 0 & 0 & 0 & 0 & 0 & 1 & 0 & 0 \\
		0 & 0 & 0 & 0 & 0 & 0 & 0 & 0 & 1 & 0 & 0 & 0 \\
		0 & 0 & 0 & 0 & 0 & 0 & 0 & 0 & 0 & 0 & 1 & 0 \\
		0 & 0 & 0 & 0 & 0 & 0 & 0 & 0 & 0 & 0 & 0 & 1 \\
		0 & 0 & 0 & 0 & 1 & 0 & 0 & 0 & 0 & 0 & 0 & 0 \\
		0 & 0 & 0 & 0 & 0 & 1 & 0 & 0 & 0 & 0 & 0 & 0 \\
		0 & 0 & 0 & 0 & 0 & 0 & 0 & 1 & 0 & 0 & 0 & 0 \\
		0 & 0 & 0 & 0 & 0 & 0 & 1 & 0 & 0 & 0 & 0 & 0 \\
		0 & 1 & 0 & 0 & 0 & 0 & 0 & 0 & 0 & 0 & 0 & 0 \\
		1 & 0 & 0 & 0 & 0 & 0 & 0 & 0 & 0 & 0 & 0 & 0 \\
		0 & 0 & 1 & 0 & 0 & 0 & 0 & 0 & 0 & 0 & 0 & 0 \\
		0 & 0 & 0 & 1 & 0 & 0 & 0 & 0 & 0 & 0 & 0 & 0
		\end{array}}
	\right) ,
	$$
	$$
	\pi (b) :=  \left( 
	{\begin{array}{rrrrrrrrrrrr}
		0 & 1 & 0 & 0 & 0 & 0 & 0 & 0 & 0 & 0 & 0 & 0 \\
		1 & 0 & 0 & 0 & 0 & 0 & 0 & 0 & 0 & 0 & 0 & 0 \\
		0 & 0 & 1 & 0 & 0 & 0 & 0 & 0 & 0 & 0 & 0 & 0 \\
		0 & 0 & 0 & 1 & 0 & 0 & 0 & 0 & 0 & 0 & 0 & 0 \\
		0 & 0 & 0 & 0 & 0 & 0 & 0 & 0 & 0 & 0 & 0 & 1 \\
		0 & 0 & 0 & 0 & 0 & 0 & 0 & 0 & 0 & 0 & 1 & 0 \\
		0 & 0 & 0 & 0 & 0 & 0 & 0 & 0 & 0 & 1 & 0 & 0 \\
		0 & 0 & 0 & 0 & 0 & 0 & 0 & 0 & 1 & 0 & 0 & 0 \\
		0 & 0 & 0 & 0 & 0 & 0 & 0 & 1 & 0 & 0 & 0 & 0 \\
		0 & 0 & 0 & 0 & 0 & 0 & 1 & 0 & 0 & 0 & 0 & 0 \\
		0 & 0 & 0 & 0 & 0 & 1 & 0 & 0 & 0 & 0 & 0 & 0 \\
		0 & 0 & 0 & 0 & 1 & 0 & 0 & 0 & 0 & 0 & 0 & 0
		\end{array}}
	\right) , 
	$$
	and 
	$$
	\pi (c) :=  \left(  
	{\begin{array}{rrrrrrrrrrrr}
		1 & 0 & 0 & 0 & 0 & 0 & 0 & 0 & 0 & 0 & 0 & 0 \\
		0 & 0 & 1 & 0 & 0 & 0 & 0 & 0 & 0 & 0 & 0 & 0 \\
		0 & 1 & 0 & 0 & 0 & 0 & 0 & 0 & 0 & 0 & 0 & 0 \\
		0 & 0 & 0 & 0 & 0 & 0 & 1 & 0 & 0 & 0 & 0 & 0 \\
		0 & 0 & 0 & 0 & 0 & 1 & 0 & 0 & 0 & 0 & 0 & 0 \\
		0 & 0 & 0 & 0 & 1 & 0 & 0 & 0 & 0 & 0 & 0 & 0 \\
		0 & 0 & 0 & 1 & 0 & 0 & 0 & 0 & 0 & 0 & 0 & 0 \\
		0 & 0 & 0 & 0 & 0 & 0 & 0 & 0 & 0 & 0 & 0 & 1 \\
		0 & 0 & 0 & 0 & 0 & 0 & 0 & 0 & 0 & 0 & 1 & 0 \\
		0 & 0 & 0 & 0 & 0 & 0 & 0 & 0 & 0 & 1 & 0 & 0 \\
		0 & 0 & 0 & 0 & 0 & 0 & 0 & 0 & 1 & 0 & 0 & 0 \\
		0 & 0 & 0 & 0 & 0 & 0 & 0 & 1 & 0 & 0 & 0 & 0
		\end{array}}
	\right) ,
	$$
	we obtain a representation $\pi$ of $S_4$ such that $ \Gamma $ is $\pi$-symmetric. We consider the regular representation $ \rho $ of $S_4$. We use Maple to find an orthonormal basis for $ K_{G}\left( \rho, \pi \right) $ and apply vectorization map to these basis elements. We obtain $288 \times 12$ matrix $ \Theta $ in \eqref{207}
	$$\Theta= \dfrac{1}{\sqrt{24}}\left( 
	\begin{array}{c}
	\pi(aa) \left[ \pi(abcaba) \right]^{-1} \\
	\pi(c) \left[ \pi(abcaba) \right]^{-1} \\
	\pi(b) \left[ \pi(abcaba) \right]^{-1} \\
	\vdots \\
	\pi(abcab) \left[ \pi(abcaba) \right]^{-1} \\
	\pi(abcaba) \left[ \pi(abcaba) \right]^{-1}  
	\end{array} \right) .
	$$
	Using \eqref{888} we could find 
	\begin{equation*}
	\Theta= \dfrac{1}{\sqrt{24}}\begin{pmatrix}
	\pi \left( g_{1}g_{1}^{-1} \right) \\\pi \left(g_{2}g_{1}^{-1} \right) \\ \vdots \\ \pi \left(g_{23}g_{1}^{-1}\right)  \\ \pi \left(g_{24}g_{1}^{-1}\right)	
	\end{pmatrix},
	\end{equation*}
	by using the ordered basis $ S_{4}=\{g_{1}, g_{2}, \ldots, g_{24} \} $. Note that the only difference between two matrices is the order of the blocks which depends on the ordering of the basis elements of $ K_{G}\left( \rho, \pi \right) $. Nonetheless, the matrices $ A_\rho $ and $ B_\rho $ in \eqref{207} are the same and we can easily compute and obtain that 
	$$A_\rho= \tilde{A}=\left( A+iB\right)^{-1}A, \quad  B_\rho =\tilde{B}=\left( A+iB\right)^{-1}B $$ and thus, the quotient graph $\Gamma_\rho$ is identical to $\Gamma$.
\end{example}

\section*{Acknowledgments} We thank Ram Band for e-mail communication suggesting to read \cite{ana} for the calculations related to Example \ref{ex54}. We also thank Ercan Altınışık for Maple calculations related to Example \ref{ex54}.



\begin{thebibliography}{99}	
	
	\bibitem{ana}
	\newblock R. Band, G. Berkolaiko, C. H. Joyner and W. Liu, 
	\newblock Quotients of finite-dimensional operators by symmetry representations, preprint, 
	\newblock 	arXiv:1711.00918 [math-ph].
	
	
	\bibitem{bandpargil} 
	\newblock R. Band, O. Parzanchevski and G. Ben-Shach,
	\newblock The isospectral fruits of representation theory: quantum graphs and drums,
	\newblock \emph{J. Phys. A: Math. Theor.}, \textbf{42} (2009), 175202.
	
	\bibitem{shasmi} 
	\newblock R. Band, T. Shapira and U. Smilansky, 
	\newblock Nodal domains on isospectral quantum graphs: the resolution of isospectrality?, 
	\newblock \emph{J. Phys. A: Math. Gen.}, \textbf{39} (2006), 13999--14014.
	
	\bibitem{ber}
	\newblock G. Berkolaiko, 
	\newblock An elementary introduction to quantum graphs, preprint, 
	\newblock 	arXiv:1603.07356 [math-ph].
	
	
	\bibitem{berkuc}
	\newblock G. Berkolaiko and P. Kuchment,
	\newblock \emph{Introduction to Quantum Graphs},
	\newblock Mathematical Surveys and Monographs vol 186, American Mathematical Society, Rhode Island, 2013.
	
	
	\bibitem{CARL} 
	\newblock R. Carlson,
	\newblock Inverse eigenvalue problems on directed graphs,
	\newblock \emph{Trans. Amer. Math. Soc.}, \textbf{351} (1999), 4069--4088.
	
	\bibitem{ful} 
	\newblock W. Fulton and J. Harris, 
	\newblock \emph{Representation Theory: A First Course},
	\newblock Readings in Mathematics, vol. 129, Springer-Verlag, New York, 1991. 
	
	\bibitem{gnusmi}
	\newblock S. Gnutzmann and U. Smilansky, 
	\newblock Quantum graphs: applications to quantum chaos and universal spectral statistics, 
	\newblock preprint, arXiv:nlin/0605028 [nlin.CD].
	
	\bibitem{gor} 
	\newblock S. Gordon, P. Perry and D. Schüth, 
	\newblock Isospectral and isoscattering manifolds: a survey of techniques and examples,
	\newblock in \emph{Geometry, Spectral Theory, Groups and Dynamics: Proceedings in Memory of Robert Brooks} (eds. M. Entov, Y. Pinchover and M. Sageev), Contemp. Math. vol 387, American Mathematical Society, (2005), 157--179.
	
	
	\bibitem{gorweb} 
	\newblock C. Gordon, D. Webb and S. Wolpert, 
	\newblock One cannot hear the shape of a drum,
	\newblock \emph{Bull. Am. Mat. Soc. (N.S.)}, \textbf{27} (1992), 134--138.
	
	\bibitem{gorweb2}
	\newblock C. Gordon, D. Webb and S. Wolpert, 
	\newblock Isospectral plane domains and surfaces via Riemannian orbifolds,
	\newblock \emph{Invent. Math.}, \textbf{110} (1992), 1--22. 
	
	
	\bibitem{gutsmi} 
	\newblock B. Gutkin and U. Smilansky, 
	\newblock Can one hear the shape of a graph?,
	\newblock \emph{J. Phys. A: Math. Gen.}, \textbf{31} (2001), 6061-6068.
	
	\bibitem{kac} 
	\newblock M. Kac, 
	\newblock Can one hear the shape of a drum?,
	\newblock \emph{Am. Math. Mon}, \textbf{73} (1966), 1--23.
	
	\bibitem{kos} 
	\newblock V. Kostrykin and R. Schrader, 
	\newblock Kirchhoff's rule for quantum wires,
	\newblock \emph{J. Phys. A: Math. Gen.}, \textbf{32} (1999), 595--630. 
	
	
	\bibitem{kotsmi} 
	\newblock T. Kottos and U. Smilansky, 
	\newblock Quantum chaos on graphs,
	\newblock \emph{Phys. Rev. Lett.}, \textbf{79} (1997), 4794--4797.
	
	
	\bibitem{kuc}
	\newblock P. Kuchment,
	\newblock Quantum graphs: an introduction and a brief survey, preprint,
	\newblock arXiv:0802.3442 [math-ph].
	
	
	
	\bibitem{mug} 
	\newblock D. Mugnolo, 
	\newblock \emph{Semigroup Methods for Evolution Equations on Networks},
	\newblock Understanding Complex Systems, Springer, Cham, 2014.
	
	\bibitem{parband} 
	\newblock O. Parzanchevski and R. Band, 
	\newblock Linear representations and isospectrality with boundary conditions,
	\newblock \emph{J. Geom. Anal.}, \textbf{20} (2010), 439--471.
	
	\bibitem{sha2} 
	\newblock T. Shapira and U. Smilansky, 
	\newblock Quantum graphs which sound the same,
	\newblock in \emph{NATO Science Series II: Mathematics, Physics and Chemistry} (eds. F. Khanna and D. Matrasulov), Nonlinear Dynamics and Fundamental Interactions vol 213, Springer, (2005), 17--29.
	
	
	\bibitem{sun} 
	\newblock T. Sunada, 
	\newblock Riemanninan coverings and isospectral manifolds, 
	\newblock \emph{Ann. Math.}, \textbf{121} (1985), 169--186. 
	
	
	\bibitem{von} 
	\newblock J. von Below, 
	\newblock Can one hear the shape of a network?,
	\newblock in \emph{Partial Differential Equations on Multistructures} (eds. F. Mehmeti, J. von Below and S. Nicaise), Lecture Notes in Pure and Applied Mathematics vol 219, Dekker, New York, (2001), 19--36. 
	
	
\end{thebibliography}
\end{document}